\documentclass[12pt]{article}
\usepackage{graphicx}
\begin{document}

\title{Effect of gauge boson mass on chiral symmetry breaking in QED$_{3}$}
\author{Guo-Zhu Liu$^{1,3}$ and Geng Cheng$^{2,3}$ \\
$^{1}${\small \textit{Lab of Quantum Communication and Quantum Computation, }%
}\\
{\small \textit{University of Science and Technology of China, }}\\
{\small \textit{Hefei, Anhui, 230026, P.R. China }}\\
$^{2}${\small \textit{CCAST (World Laboratory), P.O. Box 8730, Beijing
100080, P.R. China }}\\
$^{3}${\small \textit{Department of Astronomy and Applied Physics, }}\\
{\small \textit{University of Science and Technology of China, }}\\
{\small \textit{Hefei, Anhui, 230026, P.R. China}}}
\maketitle

\begin{abstract}
\baselineskip20pt In three-dimensional quantum electrodynamics (QED$_{3}$)
with massive gauge boson, we investigate the Dyson-Schwinger equation for
the fermion self-energy in the Landau gauge and find that chiral symmetry
breaking (CSB) occurs when the gauge boson mass $\xi$ is smaller than a
finite critical value $\xi_{cv}$ but is suppressed when $\xi > \xi_{cv}$. We
further show that the critical value $\xi_{cv}$ does not qualitatively
change after considering higher order corrections from the wave function
renormalization and vertex function. Based on the relation between CSB and
the gauge boson mass $\xi$, we give a field theoretical description of the
competing antiferromagnetic and superconducting orders and, in particular,
the coexistence of these two orders in high temperature superconductors.
When the gauge boson mass $\xi$ is generated via instanton effect in a
compact QED$_{3}$ of massless fermions, our result shows that CSB coexists
with instanton effect in a wide region of $\xi$, which can be used to study
the confinement-deconfinement phase transition.
\end{abstract}

\newpage
\baselineskip20pt

\section{INTRODUCTION}

Chiral symmetry breaking (CSB) has been an active research field in particle
physics for over forty years since Nambu [1] used this idea to generate
fermion mass in a four-fermion model. One fascinating characteristic of CSB
is that it can generate fermion mass via the fermion-antifermion
condensation mediated by strong gauge field without introducing additional
Higgs particles which till now have not been found. The most conclusive
evidence for the existence of CSB is provided by the phenomenology of strong
interaction, and CSB is widely believed to account for the pions. However,
despite the vast amount of theoretical work on CSB, till now it is not clear
whether it can be derived from quantum chromodynamics (QCD), primarily due
to the complex structure of the SU(3) gauge field. To gain valuable insight
of CSB before we can treat it completely, it is very suggestive to study
some model that is similar to QCD while is simpler than it.
Three-dimensional quantum electrodynamics (QED$_{3}$) is just such a model,
and it has attracted intense investigations [2-9] in the past twenty years.
QED$_{3}$ was shown to exhibit CSB [5-9] and confinement [9], while at the
same it is simple enough to be treated with high accuracy. Besides, it has
been used to model the physics of many planar condensed matter systems such
as high temperature superconductors [10-18] and fractional quantum Hall
systems [19].

The breakthrough in the research of CSB in QED$_{3}$ was brought by a paper
of Appelquist $et$ $al.$ [5] who found that CSB occurs when the flavor of
massless fermions is less than a critical number $N_{c}$. They arrived at
this conclusion by analytically and numerically solving the Dyson-Schwinger
(DS) equation of the fermion self-energy to the lowest order of $1/N$
expansion. Later, extensive analytical and numerical investigations [6-8]
showed that the nature of CSB in QED$_{3}$ remain the same after including
higher order corrections to the DS equation.

The above result holds when the gauge boson is massless and but is expected
to change when the gauge boson has a finite mass. CSB is a low-energy
phenomenon because (2+1)-dimensional U(1) gauge field theory is
asymptotically free [20] and only in the infrared region the gauge
interaction is strong enough to cause fermion condensation. This requires
the fermions be apart from each other. However, when the gauge boson has a
finite mass it can not mediate a long-range interaction. Intuitively, a
finite gauge boson mass is repulsive to CSB which is achieved by the
formation of fermion-anti-fermion pairs. Thus it is very interesting to
study whether CSB can occur in the presence of a finite gauge boson mass.

CSB is believed to be a nonperturbative phenomenon and hence calculations
based on perturbative expansions are incapable of establishing its
existence. We will study CSB by means of solving the nonlinear DS equation
for the fermion self-energy. Assuming that $A(p^{2})=1$ based on naive $1/N$
expansion, we get a single integral equation of the gap function $\Sigma
(p^{2})$. CSB is signalled by the appearance of a squarely integrable
nontrivial solution. To solve the DS equation, we will use bifurcation
theory and parameter imbedding method, which not only avoids the convergency
problem that usually appears in iteration method but also can help us
distinguish the different bifurcations points. After solving the DS
equation, we find that the massless fermions can acquire a finite
dynamically generated mass when the gauge boson mass $\xi $ is smaller than
a critical value $\xi _{c}$. To testify the robustness of our result against
the effect of $A(p^{2})$, we will work in a non-local gauge in which $%
A(p^{2})\equiv 1$ and the vertex function can be replaced by the gamma
matrices safely. We will show that when the wave function renormalization $%
A(p^{2})$ is included, the result we derive in the Landau gauge remains
qualitatively unchanged.

Our study on the influence of gauge boson mass on the fate of CSB not only
is of theoretical interests but can be used to understand important physical
phenomena. Actually, starting from the concept of spin-charge separation
proposed by Anderson [21], the effective low energy theory of high
temperature superconductors is a U(1) gauge theory [10-13].
Superconductivity is achieved when the charge carrying holons Bose
condensate into a macroscopic quantum state, which generate a finite mass to
the gauge boson via Anderson-Higgs mechanism. The low energy spin
fluctuations are captured by the two-component fermions, which are
originally massless since they are excited from the $d$-wave gap nodes [11].
On the other hand, CSB is known to correspond to the long-range
antiferromagnetic (AF) order, which can be seen from the behavior of AF spin
correlation function at low momentum. If we use CSB to describe the AF order
and use the gauge boson mass to describe the superconducting (SC) order, our
result then leads to a competition between the long-range AF order and the
long-range SC order, which is one of the most fundamental issues in modern
condensed matter physics. As a compromise of this competition, when the mass
of gauge boson is less than its critical value $\xi_{c}$ but is finite there
is a coexistence of these two orders in the bulk superconductors.

If the U(1) gauge field is compact in the meaning that the vector potential
has a periodicity, then it acquires a finite mass via the instanton effect.
Furthermore, permanent confinement of static charges is present when the
instanton effect is important. The influence of additional matter fields,
especially massless fermions, on the permanent confinement is an unsolved
problem. The relation between CSB and the gauge boson mass obtained in this
paper is very helpful in studying the confinement to deconfinement phase
transition driven by the couling of massless fermions.

The physical applications of CSB in the presence of a gauge boson mass to
high temperature superconductors have been reported in a Letter [16]. In
this paper, we provide the related field theoretical technique in details.
In Sec. II, we derive the DS equation in the presence of a finite mass of
the gauge boson in the Landau gauge. We then choose to solve the nonlinear
DS equation by means of bifurcation theory and parameter imbedding method.
Sec. III is devoted to the elementary knowledge of bifurcation theory that
will be used in this paper and the detailed calculation steps of parameter
imbedding method. In Sec. IV, we consider the higher order corrections to
the wave function renormalization and show that these corrections do not
change our result noticeably. In Sec.V, we give a thorough discussion of the
competing orders in high temperature superconductors from a field
theoretical point of view. In particular, we emphasize the necessity for
nonperturbative effect in getting an AF spin correlation that is consistent
with experiments. In Sec.VI, we discuss the instanton effect on CSB in
compact QED$_{3}$. The calculation of AF spin correlation function in the
CSB phase is given in the Appendix.

\section{DYSON-SCHWINGER EQUATION IN THE LANDAU GAUGE}

The three-dimentional U(1) gauge theory of massless fermions is
\begin{equation}
\mathcal{L}=\frac{1}{4}F_{\mu \nu }^{2}+\sum_{\sigma=1}^{N}\overline {\psi}%
_{\sigma} \left( \partial_{\mu }-ia_{\mu} \right)\gamma_{\mu}\psi_{\sigma},
\end{equation}
where the fermi field $\psi_{\sigma}$ is a $4\times 1$ spinor. The $4 \times
4$ $\gamma _{\mu}$ matrices obey the algebra, $\lbrace
\gamma_{\mu},\gamma_{\nu} \rbrace=2\delta_{\mu \nu}$.

The full fermion propagator is
\begin{equation}
G^{-1}(p)=i\gamma \cdot p A \left( p^{2} \right)+\Sigma \left( p^{2} \right),
\end{equation}
where $A(p^{2})$ is the wave-function renormalization and $\Sigma(p^{2})$
the fermion self-energy. The DS equation for the full fermion propagator in
momentum space is given by
\begin{equation}
G^{-1}(p)=G_{0}^{-1}(p)-\int \frac{d^{3}k}{(2\pi )^{3}}\gamma
_{\mu}G(k)\Gamma _{\nu }(p,k)D_{\mu \nu }(p-k),
\end{equation}
where $\Gamma _{\nu }(p,k)$ is the full vertex function and $D_{\mu\nu}(p-k)$
is the full photon propagator. $G_{0}^{-1}(p)$ is the bare propagator of the
massless fermions. Substituting the propagator (2) into Eq.(3) and taking
trace on both sides, we obtain the equation for $\Sigma(p^{2})$
\begin{equation}
\Sigma(p^{2})=-\frac{1}{4}\int \frac{d^{3}k}{(2\pi )^{3}}Tr\lbrack
\gamma_{\mu }G(k)\Gamma _{\nu }(p,k)D_{\mu \nu }(p-k)\rbrack.
\end{equation}
Multiplying both sides of Eq.(3) by $\gamma \cdot p$ and then taking trace
on both sides, we obtain the equation for $A(p^{2})$
\begin{equation}
A(p^{2})=1+\frac{1}{4p^{2}}\int \frac{d^{3}k}{(2\pi )^{3}}Tr\lbrack(i\gamma
\cdot k)\gamma _{\mu }G(k)\Gamma _{\nu }(p,k)D_{\mu \nu}(p-k)\rbrack.
\end{equation}

If the DS equation for $\Sigma(p^{2})$ has only vanishing solutions, the
fermions remain massless and the Lagrangian (1) respects the chiral
symmetries $\psi \rightarrow \exp (i\theta \gamma _{3,5})\psi $, with $%
\gamma _{3}$ and $\gamma _{5}$ two $4 \times 4$ matrices that anticommute
with $\gamma _{\mu}$ ($\mu=0,1,2$). If the DS equation for $\Sigma(p^{2})$
develops a squarely integrable nontrivial solution [23-25], then the
originally massless fermions acquire a finite dynamically generated mass
which breaks the chiral symmetries.

We starts from a general gauge boson propagator
\begin{equation}
D_{\mu\nu}(q)=D_{T}(q^{2})\left(\delta_{\mu\nu}-g(q^{2})\frac{q_{\mu}q_{\nu}%
}{q^{2}}\right)
\end{equation}
with
\begin{equation}
D_{T}^{-1}(q^{2})=q^{2}\left[1+\pi(q^{2})\right]+\xi^{2}
\end{equation}
where $g(q^{2})$ is a gauge-fixing parameter that depends on 3-momentum and $%
\xi$ is the mass of the gauge boson. We use $q$ to denote the gauge boson
momentum, and we have $q^{2}=(p-k)^{2}=p^{2}+k^{2}-2pk\cos{\theta}$. $%
\pi(q^{2})$ is the vacuum polarization of the gauge boson, which was
included initially to overcome the infrared divergence. If we includ only
the one-loop diagrams for massless fermions, we can write the vacuum
polarization as
\begin{equation}
\pi(q^{2})=\frac{N}{8\left|q\right|}.
\end{equation}
Then we have
\begin{eqnarray}
&D_{T}^{-1}(q^{2})&=q^{2}\pi(q^{2})+\xi^{2}  \nonumber \\
&&=\frac{N}{8}(q+\eta),
\end{eqnarray}
with
\begin{equation}
\eta=8\xi^{2}/N,
\end{equation}
since at low momentum $\pi(q^{2})\gg 1$ [5].

As the lowest-order approximation, we neglect the wave function
renormalization $A(p^{2})$ and adopt a massive gauge boson propagator in the
Landau gauge ($g(q^{2})=1$) as follows
\begin{equation}
D_{\mu\nu}(p-k)=\frac{8}{N(\left|p-k\right|+\eta)}\left( \delta_{\mu\nu}-%
\frac{(p-k)_{\mu}(p-k)_{\nu}}{(p-k)^{2}}\right).
\end{equation}
Further, we use the bare vertex, i.e., $\Gamma_{\mu}(p,k)=\gamma_{\mu}$,
which is usally called quenched planar approximation. Then the DS equation
becomes
\begin{equation}
\Sigma(p^{2})=\int\frac{d^{3}k}{(2\pi)^{3}}\frac{\gamma^{\mu}D_{\mu\nu}(p-k)
\Sigma(k^{2})\gamma^{\nu}}{k^{2}+\Sigma^{2}(k^{2})}.
\end{equation}
Now we can insert the propagator (11) into the DS equation (11), then
\begin{equation}
\Sigma(p^{2})=\frac{4}{N\pi^{2}}\int dk\frac{k\Sigma(k^{2})}{%
k^{2}+\Sigma^{2}(k^{2})}\int^{1}_{-1}dz\frac{1}{\left|p-k\right|+\eta}
\end{equation}
where $z=\cos \theta$. After performing the integration with respect to $z$
and introducing an ultraviolet cutoff $\Lambda$ we finally arrive at the
following DS equation
\begin{eqnarray}
\Sigma(p^{2})&=&\lambda\int^{\Lambda}_{0} dk\frac{k\Sigma(k^{2})}{%
k^{2}+\Sigma^{2}(k^{2})}  \nonumber \\
&&\times \frac{1}{p}\left(p+k-\left|p-k\right|-\eta\ln\left(\frac{p+k+\eta}{%
\left|p-k\right| +\eta}\right)\right),
\end{eqnarray}
where $\lambda=4/N\pi^{2}$ serves as an effective coupling constant.

If we do not introduce an ultraviolet cutoff ($\Lambda \rightarrow \infty$),
the critical behavior of Eq.(14) is completely independent of $\eta$, as can
be easily seen by making the scale transformation, $p \rightarrow p/\eta$, $%
k \rightarrow k/\eta$ and $\Sigma \rightarrow \Sigma/\eta$. We can destroy
this scale invariance by introducing an ultraviolet cutoff $\Lambda$.

Before we go into the techniques of dealing with the DS equation, we would
like to discuss one subtle issue. If the DS equation has only trivial
solutions, the fermions remain massless and the chiral symmetries are not
broken. However, $not$ all nontrivial solutions lead to CSB. It is well
known that the breaking of a chiral symmetry is always accompanied by a
Goldstone boson, which is a pseudoscalar bound state composed of a fermion
and an antifermion. If CSB happens, there should be a nontrivial solution
for the Bethe-Salpeter (BS) equation of this bound state. In addition, the
bound state wave function must satisfy a normalization condition, which can
be converted to a sufficient and necessary condition [23-25] for the
nontrivial solutions of the DS equation to signal CSB. It gives a constraint
on the form of $\Sigma(p^{2})/A(p^{2})$ as follows [25]
\begin{equation}
\int_{0}^{\infty}dq\frac{q^{2}\Sigma^{2}(q^{2})}{q^{2}A^{2}(q^{2})+%
\Sigma^{2}(q^{2})}=finite.
\end{equation}
It is easy to see that in order to satisfy this condition $
\Sigma(p^{2})/A(p^{2})$ must damp more rapidly than $p^{-1/2}$ in the
ultraviolet region ($p \rightarrow \infty$). The mass function obtained by
Appelquist $et$ $al.$ [4] satisfies this condition and hence the nontrivial
solutions of the DS equation in QED$_{3}$ corresponds to true CSB solutions
[25]. On the other hand, when an ultraviolet cutoff is introduced, the
solution $\Sigma(p^{2})$ automatically satisfies the squarely integrable
condition. We should emphasize that although the nontrivial solutions with
an ultraviolet cutoff all satisfy such a condition, only those solutions
that satisfy this condition in the continuum limit is taken are physically
sensible. In the case of four-dimensional QED, although the nontrivial
solutions with explicit ultraviolet cutoff are squarely integrable they do
not satisfy the squarely integrable condition when we take the continuum
limit [23,24]. Therefore, the CSB solutions obtained in quenched planar QED$%
_{4}$ [26] are not physically meaningful solutions because they are not
squarely integrable in the continuum limit, or in other words the associated
bound state wave functions can not be normalized. In QED$_{3}$, it was found
that $\Sigma(p^{2})/A(p^{2})$ behaves like $p^{-2}$ at $p \rightarrow \infty$%
, which surely satisfies the condition. Since the nontrivial solutions in QED%
$_{3}$ are true CSB solutions, we can safely introduce an ultraviolet cutoff
$\Lambda$ without bringing unphysical nontrivial solutions.

Theoretical analysis implies that the critical fermion number $N_{c}$ of
Eq.(4) should depend on $\Lambda/\eta$. To determine when CSB occurs, the DS
equation should be solved implicitly. The DS equation is an nonlinear
integral equation and hence is very hard to investigate. However, based on
general bifurcation theory and parameter imbedding method, we can find the
critical fermion number and the mass function exactly. The detailed programe
of calculations is the topic of the next Section. In the rest of this
Section, we discuss some qualitative properties extracted from the DS
euqation (14).

When the gauge boson has a very large mass, for example $\eta\gg\Lambda$,
then the DS equation becomes
\begin{equation}
\Sigma(p^{2})=\frac{8}{N\pi^{2}\eta}\int_{0}^{\Sigma} dk\frac{k\Sigma(k^{2})%
}{k^{2}+\Sigma^{2}(k^{2})}.
\end{equation}
From the momentum dependence of the mass function of fermions, we know that
actually it is a constant in this limit. Therefore, the DS integral equation
simplifies to a algebraic equation
\begin{equation}
\Sigma\arctan\left(\frac{\Lambda}{8\Sigma}\right)=\frac{N}{8}%
\left(1-\pi^{2}\eta\right).
\end{equation}
This equation has no solutions, hence a large enough mass of gauge boson
prevents the occurrence of CSB. We now consider another limit, i.e., when
the gauge boson mass is very small. In this limit, the last term in the
kernel of (14) can be dropped safely, leaving a DS equation that is the same
as the one studied by Appelquist $et$ $al.$ [5]. Then a very small gauge
boson mass actually does not affect the critical behavior of QED$_{3}$. This
phenomenon can be understood if it happens that the critical fermion number $%
N_{c}$ decreases when the gauge boson mass $\xi$ increases and it finally
approaches zero for very large $\xi$. The main purpose of our work is to use
QED$_{3}$ to model condensed matter systems where the physical fermion
number is $2$, which comes from the two components of the spin. Based on the
tendency of the critical fermion number in the presence of a finite gauge
boson mass, it is quite reasonable to hypothesize that there is a critical
value for the gauge boson mass $\xi_{cv}$ above which CSB is inhibited. To
make sure that it is actually the case, we should solve the DS equation and
find the critical coupling constant $\lambda_{c}$ at which the DS equation
starts to have nontrivial solutions.

\section{SOLVING DS EQUATION USING BIFURCATION THEORY AND PARAMETER
IMBEDDING METHOD}

The equation (14) is a Hammerstein type nonlinear integral equation. It does
not satisfy the conditions of the global eigenfunction theory of nonlinear
functional analysis, so its global solutions can not be obtained directly.
However, the local bifurcation theory [27] can help us to find its complete
solutions by first obtaining a local solution near a bifurcation point and
then extending its region of validity step by step. This program is most
easily achieved by parameter imbedding method [28-30], which has proved to
be a powerful method in studying integral equations.

In order to obtain the bifurcation points we need only to find the
eigenvalues of the associated Fr\^{e}chet derivative of the nonlinear DS
equation [28,29]. Those eigenvalues that have odd multiplicity are the
bifurcation points. Making Fr\^{e}chet derivative of the nonlinear equation
(14), we have the following linearized equation
\begin{equation}
\Sigma(p^{2})=\lambda\int^{\Lambda/\eta}_{0} dk\Sigma(k^{2})K(p,k)
\end{equation}
with the kernel
\begin{equation}
K(p,k)=\frac{1}{pk}\left(p+k-\left|p-k\right|-\ln\left(\frac{p+k+1}{%
\left|p-k\right|+ 1}\right)\right)
\end{equation}
where for calculational convenience we made the transformation $p
\rightarrow p/\eta$, $k \rightarrow k/\eta$ and $\Sigma \rightarrow
\Sigma/\eta$. The smallest eigenvalue $\lambda_{c}$ of this equation is just
the bifurcation point from which a nontrivial solution of the DS equation
(14) branches off. The complex kernel $K(p,k)$ in the linearized equation
(18) makes it very difficult to find an analytical solution.

We now would like to use parameter imbedding method [28,29] to solve (18)
numerically. To do this, we first analytically continue it in the complex
plane of $\lambda$, correspondingly $\Sigma(p^{2})$ also becomes a complex
function. It can be shown that
\begin{equation}
\int\int\left|K(x,y)\right|^{2}dxdy < \infty.
\end{equation}
Here we use $x$ to denote $p^{2}$, and $y$ to denote $k^{2}$. From the
Fredholm integral equation theory we know that there exists a resolvent
function for the kernel $K(x,y)$
\begin{equation}
R_{F}(x,y,\lambda)=D_{F}(x,y,\lambda)/d_{F}(\lambda),
\end{equation}
where the functions $D_{F}(x,y,\lambda)$ and $d_{F}(\lambda)$ are analytic
with respect to $\lambda$. If $d_{F}(\lambda) \neq 0$, we do not have
bifurcations points. The values of $\lambda$ at which $d_{F}(\lambda)=0$ are
the bifurcation points. According to the parameter imbedding method, the
functions $D_{F}(x,y,\lambda)$ and $d_{F}(\lambda)$ are related by the
differential-integral equations
\begin{equation}
\frac{d}{d\lambda}d_{F}(\lambda)=-\int^{\frac{\Lambda^{2}}{\eta}%
}_{0}D_{F}(x,y,\lambda)dx,
\end{equation}
\begin{equation}
\frac{\partial}{\partial\lambda}D_{F}(x,y,\lambda)=\frac{1}{d_{F}(\lambda)}%
\left[D_{F}(x,y,\lambda) \frac{d}{d\lambda}d_{F}(\lambda)+\int^{\frac{%
\Lambda^{2}}{\eta}}_{0} D_{F}(x,z,\lambda)D_{F}(z,y,\lambda)dz\right],
\end{equation}
with the initial conditions
\begin{equation}
d_{F}(0)=1,
\end{equation}
\begin{equation}
D_{F}(x,y,0)=K(x,y).
\end{equation}
One remarkable advantage of parameter imbedding method is to convert the
integral equations with variables $x$ and $y$ to a set of equations in the
variable $\lambda$. Correspondingly, the boundary conditions in original
equations are replaced by two initial conditions, which are easier to treat
in performing numerical calculations. Now, the functions $D_{F}(x,y,\lambda)$
and $d_{F}(\lambda)$ can be readily obtained by integrating numerically with
respect to $\lambda$.

We now should choose an appropriate contour $C$ in the complex $\lambda$%
-plane which contains the minimum $\lambda$ on the real axis at which $%
d(\lambda)=0$. The number of zero eigenvalues of the linearized equation
(18) inside the contour $C$ is (i.e., the zeroes of the function $%
d_{F}(\lambda)$)
\begin{equation}
N_{E}=\frac{1}{2\pi i}\oint_{C}\frac{1}{d_{F}(\lambda)}\frac{d}{d\lambda}%
d_{F}(\lambda)d\lambda.
\end{equation}
We can obtain the eigenvalues by solving the equations
\begin{equation}
\sum^{N_{E}}_{i=1}\lambda^{l}_{i}=\frac{1}{2\pi i}\oint_{C}\frac{\lambda^{l}%
}{d_{F}(\lambda)}\frac{d}{d\lambda}d_{F}(\lambda)d\lambda,
\end{equation}
with $l=1,...,N_{E}$. For the present purpose, we only need to know the
first bifurcation point, hence we let $N_{E}=1$.

For $\lambda > \lambda_{c}$, the DS equation has nontrivial solutions and
the massless fermions become massive. The ultraviolet cutoff $\Lambda$ is
provided by the lattice constant and hence is kept fixed. We can obtain the
relation of $N_{c}$ and $\eta$ by calculating the critical coupling $%
\lambda_{c}$ for different values of $\Lambda/\eta$.

Our numerical result is presented in Fig.(1). The critical fermion number $%
N_{c}$ is a monotonously increasing function of $\Lambda/\eta$. For small $%
\Lambda/\eta$, $N_{c}$ is smaller than physical number 2; so CSB does not
occur. When $\Lambda/\eta$ increases, the critical number $N_{c}$ increases
accordingly and finally becomes larger than 2 at about $\Lambda/\eta_{cv}=100
$. Then we see that there is a critical value of the gauge boson mass $%
\xi_{cv}$, below which a finite mass is generated for massless fermions
while beyond which CSB is suppressed.

\section{DYSON-SCHWINGER EQUATION WITH HIGHER ORDER CORRECTIONS}

In the last two Sections, we have investigated the DS equation in the Landau
gauge after assuming that $A(p^{2})=1$ to simplify calculations. Although
this assumption is qualitatively correct, higher order corrections from the
wave function renormalization $A(p^{2})$ will alter the critical fermion
number $N_{c}$ quantitatively. However, including $A(p^{2})$ makes the DS
equations very complicated and we should solve consistently two couples of
nonlinear integral equations. Furthermore, according to the Ward-Takahashi
identity, we can not choose $\gamma_{\mu}$ as the vertex function in the
presence of wave function renormalization $A(p^{2})$. At present, there is
no theoretical guidance in determining the vertex function $\Gamma_{\mu}(p,k)
$, and hence one can not give a guarantee of the legitimacy of a specific
choice of vertex function. Here, to simply calculations and partly overcome
the embarrassment in choosing the vertex function, we introduce a so-called
nonlocal gauge [31-34] in which the wave function renormalization $A(p^{2})
\equiv 1$ and the vertex function can be chosen as
\begin{equation}
\Gamma_{\mu}(p,k)=\gamma_{\mu}f\left(p^{2},k^{2}\right)
\end{equation}
with $f$ a function of the fermion momentum $p^{2}$, $k^{2}$. The nonlocal
gauge is obtained by solving a differential equation. In this gauge, we need
only to investigate a single equation for $\Sigma (p^{2})$ in studying the
chiral phase transition.

Let us go back to the general massive gauge boson propagator (6). If we
consider quenched planar approximation of QED$_{3}$, i.e., taking $%
\Pi(p^{2})=0$, then the wave function renormalization $A(p^{2}) \equiv 1$ in
the Landau gauge. This result is well-known to be exact in QED of dimensions
higher than $2$. In the case of QED$_{3}$, the one-loop vacuum polarization
is usually introduced explicitly to overcome the severe infrared divergence.
In the presence of $\Pi(p^{2})$, wave function renormalization $A(p^{2})$
does not equal to identity. It should be obtained by solving two consistent
integral equations of $A(p^{2})$ and $\Sigma (p^{2})$. However, taking
advantage of the gauge degrees of freedom of the system, we can simplify the
DS equations by choosing an appropriate gauge. In particular, if we could
obtain a gauge parameter $g(q^{2})$ that satisfies the following equation
[34]
\begin{equation}
g(q^{2})=\frac{2}{q^{4}D_{T}(q^{2})}\int_{0}^{q^{2}}D_{T}(z)zdz-1,
\end{equation}
then we find a gauge in which $A(p^{2})\equiv 1$. Further, according to the
Ward-Takahashi (WT) identity, the vertex function can be chosen as [34]
\begin{equation}
\Gamma_{\mu}(p,k)=\gamma_{\mu}.
\end{equation}
Now the formidable task to solve a pair of integral equations for the wave
function renormalization $A(p^{2})$ and the mass function $\Sigma(p^{2})$ is
simplified to solve a single equation of $\Sigma(p^{2})$
\begin{equation}
\Sigma(p^{2})=\int\frac{d^{3}k}{(2\pi)^{3}}\frac{\Sigma(k^{2})}{k^{2}
+\Sigma^{2}(k^{2})}\left[3-g(q^{2})\right]D_{T}(q^{2}).
\end{equation}

From $D_{T}(q^{2})$ and Eq.(29), the integral upon $z$ can be calculated
\begin{eqnarray}
\int_{0}^{q^{2}}D_{T}(z)zdz&=&\frac{8}{N}\int_{0}^{q^{2}}\frac{1}{z^{\frac{1%
}{2}} +\eta}zdz  \nonumber \\
&&=\frac{16}{N}\left(\frac{1}{3}q^{3}-\frac{1}{2}\eta
q^{2}+\eta^{2}q-\eta^{3}\ln\left(\frac{q+\eta}{\eta}\right)\right).
\end{eqnarray}
Then we obtain a nonlocal gauge parameter
\begin{equation}
g(q^{2})=\frac{4}{q^{4}}(q+\eta)\left(\frac{1}{3}q^{3} -\frac{\eta}{2}%
q^{2}+\eta^{2}q-\eta^{3}\ln\left(1+\frac{q}{\eta}\right)\right)-1.
\end{equation}
Substituting this $g(q^{2})$ into the DS equation (31), after angular
integration we have
\begin{eqnarray}
\Sigma(p^{2})&=&\frac{8}{N\pi^{2}p}\int_{0}^{\Lambda} dk\frac{k\Sigma(k^{2})%
}{k^{2}+\Sigma^{2}(k^{2})}  \nonumber \\
&&\times \int^{p+k}_{\left|p-k\right|}dq\left(\frac{2} {3}-\frac{\eta}{q+\eta%
}+\frac{\eta}{2q}-\frac{\eta^{2}}{q^{2}} +\frac{\eta^{3}}{q^{3}}\ln\left(1+%
\frac{q}{\eta}\right)\right).
\end{eqnarray}
In deriving this result, we have used the following formula
\begin{equation}
\int_{0}^{\pi}d\theta\sin{\theta}f(q^{2})=\frac{1}{pk}\int_{\left|p-k%
\right|}^{p+k}qdqf(q^{2}).
\end{equation}

After integrating (34), we have
\begin{equation}
\Sigma(p^{2})=\lambda\int^{\Lambda}_{0}dk\frac{k\Sigma(k^{2})}{k^{2}
+\Sigma^{2}(k^{2})}\frac{2}{p}K(p,k,\eta),
\end{equation}
with
\begin{eqnarray}
K(p,k,\eta)&=&\frac{2}{3}(p+k-\left|p-k\right|)-\frac{\eta}{2}\ln\left(\frac{%
p+k+\eta}{\left|p-k\right| +\eta}\right)  \nonumber \\
&&+\frac{\eta^{2}}{2}\left(\frac{1}{p+k}-\frac{1}{\left|p-k\right|}\right)
\nonumber \\
&&+\frac{\eta^{3}}{2\left|p-k\right|^{2}}\ln\left(1+\frac{\left|p-k\right|}{%
\eta}\right) -\frac{\eta^{3}}{2(p+k)^{2}}\ln\left(1+\frac{p+k}{\eta}\right).
\end{eqnarray}
Here, $\lambda=4/N\pi^{2}$ is the effective coupling constant. At the first
glance, both the third and fourth terms of $K(p,k,\eta)$ have singular
behaviors like $1/\left|p-k\right|$ which would cause divergence if $k$
approaches $p$. However, when $\left|p-k\right|\rightarrow 0$, we can make
the expansion
\begin{eqnarray}
\frac{\eta^{3}}{2\left|p-k\right|^{2}}\ln\left(1+\frac{\left|p-k\right|}{\eta%
}\right)&=&\frac {\eta^{3}}{2\left|p-k\right|^{2}}\left(\frac{%
\left|p-k\right|}{\eta}-\frac{(p-k)^{2}}{2\eta^{2}} +O\left(\left|p-k%
\right|^{3}\right)\right)  \nonumber \\
&&=\frac{\eta^{2}}{2\left|p-k\right|}-\frac{\eta}{4}+O(\left|p-k\right|).
\end{eqnarray}
Thus the singular terms are exactly cancelled. The same step can be used to
show that the singular term $1/(p+q)$ can also be cancelled exactly.
Therefore, the kernel $K(p,k,\eta)$ is a smooth function on the whole
integration region.

Making Fr\^{e}chet derivative of the nonlinear equation (35), we obtain the
linearized equation
\begin{equation}
\Sigma(p^{2})=\lambda\int^{\frac{\Lambda}{\eta}}_{0}dk\Sigma(k^{2})\frac{2}{%
pk}K(p,k,\eta)
\end{equation}
with
\begin{eqnarray}
K(p,k,\eta)&=&\frac{2}{3}(p+k-\left|p-k\right|)-\frac{1}{2} \ln\left(\frac{%
p+k+1}{\left|p-k\right|+1}\right)  \nonumber \\
&&+\frac{1}{2}\left(\frac{1}{p+k}-\frac{1}{\left|p-k\right|}\right)
\nonumber \\
&&+\frac{1}{2\left|p-k\right|^{2}}\ln\left(1+\left|p-k\right|\right) -\frac{1%
}{2(p+k)^{2}}\ln\left(1+p+k\right)
\end{eqnarray}
where for calculational convenience we made the transformation $p
\rightarrow p/\eta$, $k \rightarrow k/\eta$ and $\Sigma\rightarrow
\Sigma/\eta$.

Using the steps we presented in the last Section, we can solve the
linearized equation (39) to obtain the relation between the critical fermion
number $N_{c}$ and the mass $\xi$ of the gauge boson mass. The numerical
result is presented in Fig.(2), from which we know that the critical value
of the gauge boson mass is about $\Lambda/\eta_{cv}=3.3$. Although there is
a significant change in the critical value $\xi_{cv}$, the result we
obtained in the Landau gauge remains qualitatively correct.

\section{COMPETING ORDERS IN HIGH TEMPERATURE SUPERCONDUCTORS}

Understanding the competing orders in high temperature cuprate
superconductors is one of the most important issues in condensed matter
physics. In the presence of competing orders, one order parameter prevails
when other orders are suppressed by some external variables. At
half-filling, the cuprate superconductor is a Mott insulator with long-range
antiferromagnetic (AF) order. When holes are dopped into the Cu-O planes,
the material becomes a superconductor at low temperatures and the long-range
AF order disappears. Hence there is a competition between the AF order and
the SC order, and as a result of this competition the AF order dominates at
zero and low doping while the SC order dominates at higher doping. However,
even at higher doping the AF order also has a chance to appear locally where
the superconductivity is suppressed by strong external magnetic fields.
Recently, elaborate neutron scattering [35] and scanning tunnelling
microscopy (STM) [36] experiments found that the AF correlation is
significantly enhanced in regions surrounding the vortex cores. In this
paper, we will use spin-charge separation and CSB to understand the
competing orders.

It has been shown that [10-13] Lagrangian (1) is the effective low energy
theory of undoped cuprates which has only fermionic excitations because of
the presence of a large charge gap. In underdoped cuprates, the electrons
fractionalize into spin carrying spinons and charge carrying holons. It has
been pointed out [15,16] that the physics of underdoped cuprates is captured
by an effective U(1) gauge theory of massless fermions and charged scalar
fields
\begin{equation}
\mathcal{L}_{F}=\sum_{\sigma=1}^{N}\overline{\psi}_{\sigma} v_{\sigma, \mu}
\left( \partial_{\mu }-ia_{\mu} \right)
\gamma_{\mu}\psi_{\sigma}+\left|\left(\partial_{\mu}-ia_{\mu}
\right)b\right|^{2}+V(\left|b\right|^{2}).
\end{equation}
Here $b=(b_{1},b_{2})$ is a doublet of scalar fields representing the holons
[15]. $v_{\sigma, 0}=1$ and generally $v_{\sigma, 1} \neq v_{\sigma, 2}$ as
a result of the velocity anisotropy; however, for simplicity we can let $%
v_{\sigma, 1}=v_{\sigma, 2}=1$. Since the spin and charge degrees of freedom
are assumed to be separated, there is no Yukawa-type coupling between the
fermion field and the scalar field. In the superconducting state, the boson $%
b$ acquires a nonzero vacuum expectation value, i.e., $\langle b \rangle
\neq 0$. This nonzero $\langle b \rangle$ spontaneously breaks gauge
symmetry of the theory and the gauge boson acquires a finite mass $\xi$ via
Anderson-Higgs mechanism.

In the context of high temperature superconductors the U(1) gauge field is
introduced as a Lagrangian multiplier to impose local no-double occupancy
constraint. It has no kinetic term $\sim F_{\mu\nu}^{2}$ and its dynamics is
obtained by integrating out the matter fields. If we only include the
one-loop diagram in the vacuum polarization, we get $D^{-1}_{T}(q^{2})=q^{2}%
\pi(q^{2})+\xi$. As we have showed previously, the effect of additional
scalar doublet is to shift $N$ in the gauge boson vacuum polarization $%
\pi(q^{2})$ to $N+1$, i.e., $\pi(q^{2})=(N+1)/8\left| q \right|$. Then the
propagator for the gauge boson is
\begin{equation}
D_{\mu\nu}(p-k)=\frac{8}{(N+1)(\left|p-k\right|+\eta)}\left( \delta_{\mu\nu}-%
\frac{(p-k)_{\mu}(p-k)_{\nu}}{(p-k)^{2}}\right).
\end{equation}
From the corresponding DS equation in the non-local gauge obtained above, we
found a critical gauge boson mass at $\Lambda/\xi_{cv}=100$. For small $\xi$%
, CSB occurs; while for $\xi > \xi_{cv}$, CSB is suppressed.

We now would like to discuss the long-range behavior of the AF correlation
function. The AF spin correlation is defined as
\begin{equation}
\langle S^{+}S^{-}\rangle _{0}=-\frac{1}{4}\int \frac{d^{3}k}{(2\pi )^{3}}Tr%
\left[ G_{0}(k)G_{0}(k+p)\right] ,
\end{equation}%
where $G_{0}(k)$ is the fermion propagator. If the fermions are massless,
then
\begin{equation}
G_{0}(k)=-\frac{\gamma \cdot k}{k^{2}},
\end{equation}%
and we have
\begin{equation}
\langle S^{+}S^{-}\rangle _{0}=-\frac{\left| p\right| }{16}.
\end{equation}%
At $p\rightarrow 0$, $\langle S^{+}S^{-}\rangle _{0}\rightarrow 0$, and the
AF correlation is heavily lost. This is not a surprising result since our
starting point is the resonating valence bond (RVB) picture proposed by
Anderson [21], which is just a liquid of spin singlets and hence it has only
short range AF correlation. This is not a satisfying situation because a
long-range N\'{e}el order was observed in experiments shortly after the
discovery of cuprate superconductors.

However, even if we starts from an RVB ground state, it is still possible to
obtain the long-range AF correlation because of the strong correlation
nature of the Mott insulators. The strong correlation is reflected in the
fact that double occupancy on a single lattice is completely inhibited due
to the strong Coulomb repulsive force. After this local constraint and
quantum fluctuations are taken into account, a strong U(1) gauge field
emerges in the effective theory. This gauge field has important effect on
physical properties since it can cause fermion condensation and give the
originally massless fermions a finite mass. The AF spin correlation is
expected to be significantly enhanced once the fermions become massive. To
show this actually happens, we will calculate the spin correlation function
in the CSB phase (see Appendix for details). Although the dynamically
generated fermion mass depends on the 3-momentum, here, for simplicity, we
assume a constant mass $m$ for the fermions. This approximation is valid
because we only care about the low-energy property and $\Sigma(p^{2})$ is
actually a constant at $p \rightarrow 0$. The propagator for the massive
fermion is
\begin{equation}
G(k)=\frac{-\left(\gamma \cdot k+im\right)}{k^{2}+m^{2}},
\end{equation}
which leads to
\begin{equation}
\langle S^{+}S^{-}\rangle_{0}=-\frac{1}{4\pi}\left(m +\frac{p^{2}+4m^{2}}{%
2\left|p\right|} \arcsin\left(\frac{p^{2}}{p^{2}+4m^{2}}\right)^{1/2}\right).
\end{equation}
This spin correlation behaves like $-m/2\pi$ as $p \rightarrow 0$ and we
have long-range AF correlation when CSB takes place. Therefore, strong
fluctuations around the RVB ground state enhances the long-range AF spin
correlations.

We should emphasize that calculations based on perturbative expansions can
not be used to obtain the long-range AF order. It might be argued that
including higher order diagram can enhance the AF spin correlation. However,
this argument is not right. If we include the gauge field while keeping the
fermions massless, then the spin correlation is [37]
\begin{equation}
\langle S^{+}S^{-}\rangle_{GF}=-\frac{8}{12\pi^{2}(N+1)}\left|p\right|\ln
\left(\frac{\Lambda^{2}}{p^{2}}\right)
\end{equation}
which damps at low momentum $p \rightarrow 0$. Rantner and Wen [37] used to
claim that long-range AF correlation can be obtained by reexponentiating the
spin correlation function [38]. This scenario is based on their previous
statement [14] that the U(1) gauge field can not generate a finite mass for
fermions and hence is a marginal perturbation. This result is derived by
considering only the one-loop correction of gauge field to the fermion
self-energy. However, CSB is a nonperturbative phenomenon and whether the
gauge field generates a finite mass for the massless fermions can only be
settled by investigating the self-consistent DS equation for the fermion
self-energy. If the equation (12) does not have the nonlinear term in the
denominator of the kernel, it is a linear equation and can $not$ develop any
genuine nontrivial solution. From the point of view of bifurcation theory, a
linear operator has no bifurcation points those are necessary for a phase
transition to take place. Once the nonperturbative effect is taken into
account, the strong gauge field generates a finite fermion mass which breaks
the chiral symmetry and gives rise to long-range AF order (Ref.[39]
discussed the correspondence of CSB to AF order in another way). Actually,
the formation of long-range AF order spontaneously breaks the rational
symmetry of the system and generates a gapless spin wave excitation which
corressponds to the massless Goldstone boson. These are hard to understand
if we only include the gauge fluctuations without breaking any symmetry.
Furthermore, the strong interaction of the gauge field with massless
fermions of flavor $2$ will unavoidably generate a finite fermion mass.

Now we would like to discuss the application of our result to the interplay
of various ground states in high temperature cuprate superconductors. It is
well-known that the gauge boson mass $\xi$ is proportional to the superfluid
density $\rho$, thus we can use $\xi$ to describe the superconducting order.
Otherwise, we use CSB to describe the long-range AF order. Based on the fact
that the superfluid density is proportional to doping concentration, we
obtain a clear picture of the evolution of different orders upon increasing
the doping concentration. At zero and low doping the gauge boson mass is
zero or very small, so CSB and hence the AF order is present. When the
doping concentration is larger than a critical value $\delta_{cv}$ the gauge
boson acquire a mass that is large enough to suppress the CSB and the AF
order. Note that superconductivity begins to appear as the ground state of
cuprate superconductors at $\delta_{sc}$ which is less than $\delta_{cv}$.
Therefore, for $\delta_{sc} < \delta < \delta_{cv}$ there is a coexistence
of the AF order and the SC order in the bulk materials. Due to this
coexistence, the length scale for AF order to appear should be larger than
the vortex scale, which is consistent with STM experiments of Hoffman $et$ $%
al.$ [36].

When the external magnetic field is stronger than $H_{c2}$, the superfluid
is completely suppressed, and, correspondingly, the gauge boson become
massless. Then CSB reappears in the bulk material and gives a mass to the
massless fermions. This mass provides a finite gap for the low energy
fermions to be excited, thus at low temperature no fermionic excitations can
be observed [15]. This causes the breakdown of the Wiedemann-Franz (WF) law
in the normal ground state of cuprate superconductors [15,40]. Thus, based
on spin-charge separation and CSB, we give a unified description for both
the behaviors of AF spin correlation and the transport properties from a
field theoretical point of view. This is the most noticeable advantage of
our scenario comparing with so many other scenarios [41-47] those also
address the problem of local AF order in vortex cores.

\section{INSTANTON EFFECT ON CSB}

Confinement is one unresolved problem in modern particle physics. A seminal
paper written by Polyakov [22] has shed some light on this problem by
studying a three-dimentional compact pure U(1) gauge theory (compact pure QED%
$_{3}$). In general, one can define an Abelian gauge theory on a
two-dimensional lattice which has the following action
\begin{equation}
S=\frac{1}{2e^{2}}\sum_{i,\alpha \beta}\left(1-\cos F_{i,\alpha
\beta}\right),
\end{equation}
with the field strength
\begin{equation}
F_{i,\alpha
\beta}=A_{i,\alpha}+A_{i+\alpha,\beta}-A_{i+\beta,\alpha}-A_{i,\beta}.
\end{equation}
Here, the pairs ($i$,$\alpha$) are used to denote the links between
lattices, with $i$ the beginning of a link and $\alpha$ its direction. If
the vector potential $A_{i,\alpha}$ is defined to be a real number on its
whole region, i.e., $-\infty \leq A_{i,\alpha} \leq +\infty$, the continuum
limit of this action is just that of the usual U(1) gauge as presented in
(1). However, a highly nontrivial physical effect emerges if the vector
potential $A_{i,\alpha}$ has angular properties and hence is defined on a
circle as $-\pi \leq A_{i,\alpha} \leq \pi$. Due to the periodicity of its
action, such a field theory is called compact QED.

Polyakov firstly considered the pure compact QED$_{3}$ without coupling
matter fields to the gauge field. He found that instantons appear in this
model as topological solutions of the Euclidean gauge field equations and
lead to permanent confinement of static charges which is reflected by the
area law for the Wilson integral. Compact QED$_{3}$ has attracted intense
investigations in the past twenty years, initially as a simpler model to
study quark confinement. Recently, compact QED$_{3}$ with matter fields has
been used to model the physics of many strongly correlated electron systems
[19,48]. However, although it is widely accepted that confinement is present
in pure compact QED$_{3}$, there is no consensus on the fate of permanent
confinement when matter fields are included [49,50]. Comparing with compact
QED$_{3}$ of scalar fields [49], the situation for compact QED$_{3}$ of
massless fermions is particularly complicated because of the possibility of
dynamical mass generation for the fermions.

Since compact QED$_{3}$ is originally defined on lattices, Monte Carlo
numerical simulations are expected to provide important information on CSB,
but they suffers from the notorious fermion sign problem. In this paper, we
would like to analyze the chiral behavior using the DS equation method. To
do this, we map the compact QED$_{3}$ onto a continuum theory and introduce
the ultraviolet cutoff $\Lambda$ keeping track of its lattice origin. As
shown by Polyakov, the gauge field acquires a finite mass due to Debye
screening caused by the instantons. We can use the mass $\xi$ of gauge field
to describe the instanton effect and investigate the relation between CSB
and instanton effect by solving the DS equation that consists of a massive
gauge boson propagator.

We have studied the relation of gauge boson mass and CSB in the context of
high temperature superconductors in the last section. However, the critical
gauge boson mass $\xi_{cv}$ is very small in the presence of additional
scalar fields, due to the shift from $N$ to $N+1$ by the scalar doublet, and
hence CSB can exist only for a small region of $\xi$. But the critical value
$\xi_{cv}$ in the present case (compact U(1) gauge field coupled only to
massless fermions) is much larger and there is a wide region of $\xi$ for
CSB to take place. In the previous papers [5,6] addressing CSB in QED$_{3}$
the ultraviolet cutoff is provided by $\alpha=N/8$, which is kept fixed when
the fermions flavor $N$ is taken to infinity, because for momentum $p>\alpha$
the self-energy function damps rapidly. From $\eta=\xi^{2}/\alpha$ we know
that the critical gauge boson mass is about $\xi_{c}=\alpha/2$. The
instanton effect can coexist with CSB for $\xi<\alpha/2$. If we couple a
fermion of one flavor to compact gauge field, then CSB can coexist with
instanton effect in a much wider region of $\xi$.

The above result can be used to investigate the possible confinement to
deconfinement transition in compact QED$_{3}$ because whether the fermions
have a finite mass is expected to affect the fate of permanent confinement
[51]. Such a transition is no doubt of great importance in both particle
physics and condensed matter physics, but beyond the scope of this paper.

\section{Summary and Discussion}

In this paper, we have discussed the effect of a finite mass of U(1) gauge
boson on CSB and its physical implications. The gauge boson mass $\xi$ is
reflected in the modification of the gauge field propagator, which appears
in the DS equation of the fermion self-energy. The DS equation is nonlinear
and hence hard to be solved. Iteration procedure is the most frequently used
numerical calculation method, but it is not clear whether the iteration
procedure leads to a convergent result or not. To avoid the problem brought
by the convergency of iteration, we make use of bifurcation theory and
parameter imbedding method to numerically investigate the DS equation.
Adopting the Landau gauge and neglecting the wave function renormalization,
we found a critical value $\xi_{cv}$ for the gauge boson mass that separates
the CSB phase, for $\xi < \xi_{cv}$, and chiral symmetric phase, for $\xi >
\xi_{cv}$. We then showed that including higher order corrections of the
wave function renormalization does not change qualitatively the critical
value $\xi_{cv}$.

We then use our result to two physical systems, high temperature cuprate
superconductors and compact QED$_{3}$. If the gauge boson mass is generated
via Anderson-Higgs mechanism in the superconducting state, the combination
of spin-charge separation and CSB provides a field theoretical description
of the competition between the AF order and the SC order. As a compromise of
this competition, there is a microscopic coexistence of these two orders in
the bulk materials, which plays an essential role in explaining the local AF
ordre in vortex states observed in neutron scattering and STM experiments.
When the periodicity of the gauge field is taken into account, the gauge
boson acquires a finite mass via the instanton effect. Since whether the
permanent confinement still exists in the presence of fermions depends on
the fermion mass, our result can help us to investigate the confinement to
deconfinement phase transition, which will be the subject of future study.

\section*{Acknowledgement}

We would like to thank Cheng Lee for his help in numerical calculations and
V. P. Gusynin for helpful communications. This work is supported by National
Science Foundation in China No.10175058.

\section*{Appendix}

In this Appendix we give the details of calculating the spin correlation
using the propagator of the massive fermions. When CSB occurs the fermion
propagator is
\begin{equation}
G(k)=\frac{-\left( \gamma \cdot k+im\right) }{k^{2}+m^{2}}.
\end{equation}%
Then
\begin{eqnarray}
&\langle S^{+}S^{-}\rangle _{0}=&-\frac{1}{4}\int \frac{d^{3}k}{(2\pi )^{3}}%
Tr\left[ G(k)G(k+p)\right]   \nonumber \\
&=&-\frac{1}{4}\int \frac{d^{3}k}{(2\pi )^{3}}Tr\left[ \frac{\gamma \cdot
k+im}{k^{2}+m^{2}}\frac{\gamma \cdot (k+p)+im}{(k+p)^{2}+m^{2}}\right]
\nonumber \\
&=&-i\int_{0}^{1}dt\int \frac{d^{3}k}{(2\pi )^{3}}\frac{k\cdot (k+p)-m^{2}}{%
\left[ \left( k^{2}+m^{2}\right) (1-t)+\left( (k+p)^{2}+m^{2}\right) t\right]
^{2}},
\end{eqnarray}%
where we used the Feynman parameterization formula
\begin{equation}
\frac{1}{ab}=\int_{0}^{1}dt\frac{1}{\left[ at+b(1-t)^{2}\right] ^{2}}.
\end{equation}%
After replacing $k$ by $k-pt$ and making Wick rotation, we have
\begin{equation}
\langle S^{+}S^{-}\rangle _{0}=\int_{0}^{1}dt\int \frac{d^{3}k}{(2\pi )^{3}}%
\frac{k^{2}-m^{2}-p^{2}t(1-t)}{\left[ k^{2}+m^{2}+p^{2}t(1-t)\right] ^{2}}.
\end{equation}%
Using the properties of the $\Gamma $-function, we can integrate upon the
momentum $k$ and get
\begin{equation}
\langle S^{+}S^{-}\rangle _{0}=\frac{1}{(4\pi )^{\frac{3}{2}}}\frac{\frac{3}{%
2}\Gamma (-\frac{1}{2})-\Gamma (\frac{1}{2})}{\Gamma (2)}\int_{0}^{1}dt\frac{%
1}{\left[ m^{2}+p^{2}t(1-t)\right] ^{-\frac{1}{2}}}.
\end{equation}%
Since
\begin{equation}
\frac{\Gamma (-\frac{1}{2})}{\Gamma (2)}=-2\pi ^{\frac{1}{2}}
\end{equation}%
\begin{equation}
\frac{\Gamma (\frac{1}{2})}{\Gamma (2)}=\pi ^{\frac{1}{2}},
\end{equation}%
we get
\begin{eqnarray}
&\langle S^{+}S^{-}\rangle _{0}=&-\frac{1}{2\pi }\int_{0}^{1}dt\left[
m^{2}+p^{2}t(1-t)\right] ^{\frac{1}{2}}  \nonumber \\
&=&-\frac{1}{4\pi }\left( m+\frac{p^{2}+4m^{2}}{2\left| p\right| }\arcsin
\left( \frac{p^{2}}{p^{2}+4m^{2}}\right) ^{\frac{1}{2}}\right) .
\end{eqnarray}%
This is the AF spin correlation in the CSB phase.

\begin{figure}
\centering \includegraphics{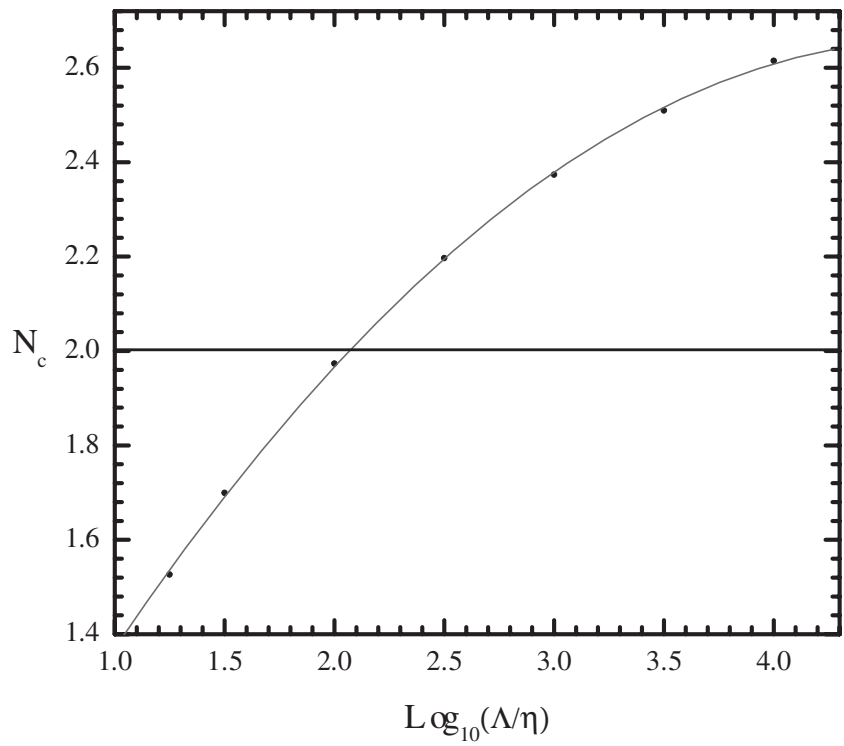}
\begin{minipage}{10cm}
\caption{The dependence of the critical number $N_{c}$ on
$\log_{10}(\Lambda/\eta)$ in the Landau gauge.}
\end{minipage}

\end{figure}

\begin{figure}
\centering \includegraphics{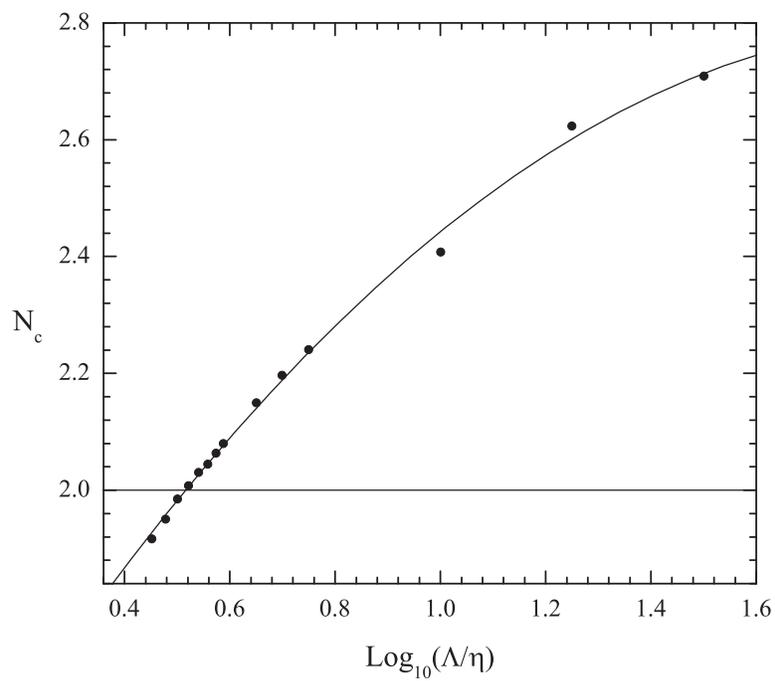}
\begin{minipage}{12cm}
\caption{The dependence of the critical number $N_{c}$ on
$\log_{10}(\Lambda/\eta)$ in the non-local gauge.}
\end{minipage}

\end{figure}

\end{document}